 \documentclass[final,1p,times]{elsarticle}

\usepackage{graphicx}
\usepackage{bm}
\usepackage{amssymb}
\usepackage{color}

\journal{Annals of Physics}

\begin{document}
\newcommand{\mvec}[2]
{
\left(\begin{array}{c}
#1  \\
#2  
\end{array}
\right)
}
\newcommand{\mvecthree}[3]
{
\left(\begin{array}{c}
#1  \\
#2  \\
#3  
\end{array}
\right)
}
\newcommand{\mvecfour}[4]
{
\left(\begin{array}{c}
#1  \\
#2  \\
#3  \\
#4  
\end{array}
\right)
}
\newcommand{\mmat}[4]
{
\left(\begin{array}{cc}
#1  & #2\\
#3  & #4
\end{array}
\right)
}

\begin{frontmatter}



\title{Robust Zero Modes in Disordered Two-Dimensional Honeycomb Lattice with Kekul\'{e} Bond Ordering}


\author[label1]{Tohru Kawarabayashi}
\author[label1]{Yuya Inoue}
\author[label1]{Ryo Itagaki}

\address[label1]{Department of Physics, Toho University, Funabashi 274-8510, Japan}
\author[label2]{Yasuhiro Hatsugai}

\address[label2]{Department of Physics, University of Tsukuba, Tsukuba 305-8571, Japan}
\author[label3,label4]{Hideo Aoki}
\address[label3]{Department of Physics, University of Tokyo, Hongo, 
Tokyo 113-0033, Japan}
\address[label4]{Electronics and Photonics Research Institute, 
Advanced Industrial Science and Technology (AIST), 
Tsukuba, Ibaraki 305-8568, Japan}
\begin{abstract}
Robustness of zero-modes of two-dimensional Dirac fermions is examined numerically for the honeycomb lattice in the 
presence of Kekul\'{e} bond ordering. The split $n=0$ Landau levels in a magnetic field as well as the zero-modes generated by 
topological defects in the Kekul\'{e} ordering 
are shown to exhibit anomalous robustness against disorder 
when the chiral symmetry is respected.

\end{abstract}




\end{frontmatter}


\section{Introduction}

Graphene \cite{Geim,Kim} has kicked off the physics of Dirac fermions in two dimensions as one of the  central issues in 
the condensed-matter physics \cite{R1,R2,R3,Ando,PG}. In particular, the characteristic quantum Hall effect is a hallmark of the graphene physics. 
In the developments of theoretical and experimental understanding of the electronic states for Dirac fermions, 
there has been a considerable progress in the theoretical studies\cite{HCM,CHJMPS,CHJMPS2,RMHC} of the charge fractionalization in two dimensions where fractional and irrational charges emerge in association with the topological 
defects (vortices) of the Kekul\'{e} bond ordering in the honeycomb lattice. 
Such a system is generally known as a fermion-vortex system having topological zero-energy state in a bulk gap 
\`{a} la Jackiw and Rossi \cite{JR}. 
Experimentally, the Kekul\'{e} bond ordering is now realized not only in molecular graphene\cite{M-Graphene} but also in photonic 
crystals \cite{YXXWJHH} in a controlled manner enabling 
us to examine their topological properties.

With such a recent progress in mind, here we 
explore how disorder will affect the topological zero-energy state associated with the 
vortex structure in the Kekul\'{e} bond ordering in two dimensions, along with the $n=0$ Landau levels split by the Kekul\'{e} ordering. 
We evaluate the density of states numerically for the lattice model by adopting the Green function method \cite{SKM} and the kernel polynomial method \cite{KPM}. 
To discuss the topological zero-modes induced by topological defects exemplified here by vortices in the Kekul\'{e} bond ordering, numerical 
calculations in large enough systems are necessary for a quantitative analysis. The kernel polynomial method is suitable for obtaining the local density of states accurately in large systems. 
We have adopted the method to examine not only a single vortex in the bond ordering but also vortices with higher winding numbers and 
a pair of the vortex and the anti-vortex.  

\section{Kekul\'{e} bond ordering}
\begin{figure}
\includegraphics[scale=0.6]{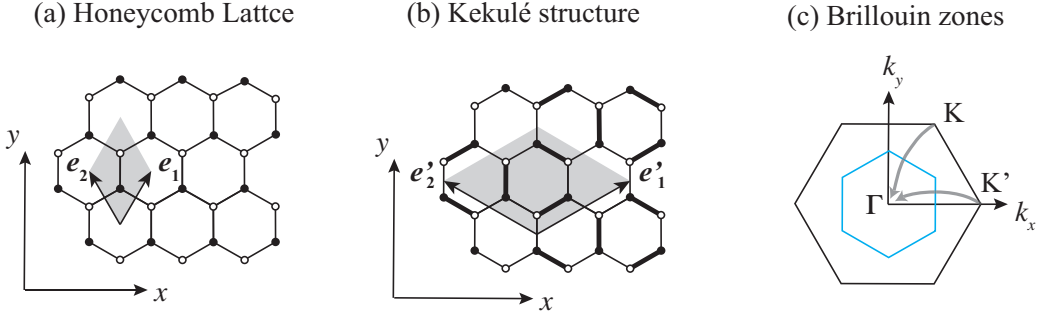}
\caption{(Color online)
(a) The usual honeycomb lattice, for which 
a unit cell is indicated by the 
primitive vectors  $\bm{e}_1=(\sqrt{3}/2, 3/2)a$ and $\bm{e}_2=(-\sqrt{3}/2,3/2)a$ with 
the distance $a$ between the nearest-neighbor sites. The filled (open) circles represent the A (B) sub-lattice. 
(b) The honeycomb lattice with the Kekul\'{e} bond 
ordering. The thick (thin) bond represents the hopping amplitude $t +2\Delta_K \; (t-\Delta_K)$.
The unit cell is now given by the primitive vectors $\bm{e}'_1= (3\sqrt{3}/2, 3/2)a$ and  $\bm{e}'_2= (-3\sqrt{3}/2, 3/2)a$. (c) The first Brillouin 
zone for the honeycomb lattice (the outer hexagon) 
is folded (arrows) onto the inner hexagon (blue) in the presence of the Kekul\'{e} bond ordering, where 
the Brillouin zone becomes 1/3 of the original one and the K and K' points folded onto $\Gamma=(0,0)$.
}
\label{honeycomb}
\end{figure}

We consider non-interacting electrons on the two-dimensional honeycomb lattice with the nearest-neighbor hopping $t$.
The energy dispersion is given by 
$
 E(\bm{k}) =  \pm |d(\bm(k)| 
$
with $d(\bm{k}) = t[1+\exp(-ik_1)+\exp(-ik_2)]$, where $k_i$ $(i=1,2)$ denotes the projection $\bm{k}\cdot \bm{e}_i $ of the wave vector on 
the primitive vectors $\bm{e}_1 = (\sqrt{3}/2, 3/2)a$ and $\bm{e}_2 = ( -\sqrt{3}/2, 3/2)a$ (Fig. \ref{honeycomb}(a)) 
with $a$ being the 
nearest-neighbor distance in the honeycomb lattice. The valence and conduction bands touch with each other 
at K and K' points, $(k_1, k_2) = (-2\pi/3, 2\pi/3)$ and $ (2\pi/3, -2\pi/3)$,  in the first 
Brillouin zone, at which $d(\bm{k})$ 
vanishes and a massless Dirac dispersion emerges.  The Kekul\'{e} bond ordering shown in Fig.
 \ref{honeycomb} (b) folds the K and K' points onto $\Gamma$ point $(0,0)$ in the Brillouin zone, at which 
an energy gap 
 $\pm 3\Delta_K$ at $\Gamma$ opens.  
$\Delta_K$ is determined by the degree of the 
Kekul\'{e} bond ordering (e.g., the difference 
in the transfer 
energy between ``thick" and ``thin" bonds 
in a tight-binding model as depicted in Fig.1).  
The massless Dirac fermions then acquire a mass 
(in zero
magnetic field), despite the fact that the bond ordering does not break the chiral symmetry.

The Kekul\'{e} bond ordering for the honeycomb lattice has been discussed as an instability in a strong magnetic field, where the $n=0$ Landau level 
splits into $\pm 3\Delta_K$ \cite{VAA,AA,HFA}. 
Roles of the topological defects in the Kekul\'{e} bond ordering have also been
discussed for transport properties of graphene in magnetic fields \cite{NRL}.   

The tight-binding Hamiltonian for the honeycomb lattice with the nearest-neighbor hopping is given as 
$$
 H = \sum_{<i,j>} t_{i,j}e^{-2\pi i\theta_{i,j}} c_i^\dagger c_j + {\rm h. c.}
$$
in standard notations, 
where the nearest-neighbor hopping $t_{i,j} = t + t_K +\delta t_{i,j}$ is real. We consider the Kekul\'{e} bond 
ordering given as $t_K = 2\Delta_K \; (-\Delta_K)$ for the thick (thin) bonds in Fig. \ref{honeycomb} (b), while 
$\delta t_{i,j}$ represents a random component 
which breaks the translational invariance of the system. 
The magnetic field perpendicular to the system is 
incorporated as the Peierls phase $\theta_{ij}$ determined so that the sum
of the phases along a closed loop is equal to the magnetic flux enclosed by the loop in units of the flux quantum $(h/e)$.  Let us denote the magnetic flux per hexagon as $\phi$.

\section{Robust $n=0$ Landau levels}
Based on the tight-binding model on the honeycomb lattice, we examine the robustness of the $n=0$ Landau levels 
in the presence of the Kekul\'{e} bond ordering 
against an introduction of disorder to the hopping amplitudes which 
breaks the translational symmetry. 
In the absence of the Kekul\'{e} bond ordering, it has been shown that the $n=0$ Landau level exhibits an anomalous sharpness 
when the bond disorder is spatially correlated over several lattice constants \cite{KHA, KMHA}, 
which can be understood as a manifestation of the single 
Dirac fermion property \cite{AC}, since the scattering between two Dirac fermions at K and K' points is strongly suppressed in such a situation.  
In the presence of the Kekul\'{e} bond ordering, 
on the other hand, the crucial question is whether or not the anomalous sharpness of the $n=0$ Landau level is degraded by the strong mixing between K and K' valleys. 

We consider a random component $\delta t_{i,j} $ that 
obeys a Gaussian probability distribution with a variance $\sigma$,
$$
 P(\delta t_{i,j}) = \frac{1}{\sqrt{2\pi \sigma}} \exp\bigg(-\frac{\delta t^2}{2\sigma}\bigg).
$$
The random components are assumed to be spatially correlated as 
$$
 \langle \delta t_{i,j} \delta t_{k,l} \rangle = \langle \delta t^2\rangle \exp(-(r_{i,j}-r_{k,l})^2/4\eta),
$$
where $\eta$ denotes the spatial correlation length and $r_{i,j}$ denotes the spatial position of the bond $t_{i,j}$.
We should note that the present disordered system 
still respects the chiral symmetry.

\begin{figure}
\includegraphics[scale=0.4]{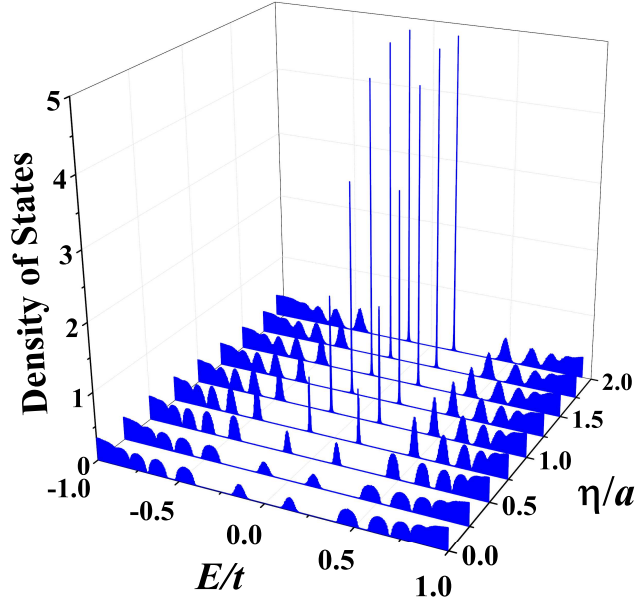}
\caption{
(Color online) Density of states $\rho(E) = -\langle {\rm Im}G_{ii}(E+i\varepsilon)\rangle/\pi$ 
for the honeycomb lattice with the Kekul\'{e} bond ordering $\Delta_K/t= 0.05$  in a uniform magnetic field 
$\phi/(h/e) = 1/50$ is plotted for various values of the spatial correlation length  $\eta$ for the random component of the hopping amplitudes. 
We assume the strength of disorder to be $\sigma /t = 
0.115$, and the imaginary part of the energy 
$\varepsilon/t = 6.25\times 10^{-4}$.
}
\label{dos}
\end{figure}

We evaluate with  the Green function method \cite{SKM} 
the averaged density of states $\rho(E) = -\langle {\rm Im}G_{ii}(E+i\varepsilon)\rangle/\pi$ with $G_{ii}(E) = \langle i|(E-H)^{-1}|i\rangle$, where the angular bracket $\langle \rangle$ denotes the average over the sites $i$.  
If we look at the result in Fig. \ref{dos}. 
the split $n=0$ Landau levels at $E=\pm 3\Delta_K$ exhibit an anomalous sharpness as soon as 
the spatial correlation length $\eta$ exceeds the nearest-neighbor distance $a$ of the honeycomb lattice. 
Thus we can conclude that the anomalous robustness 
of the $n=0$ Landau levels against the correlated bond disorder survives even in the presence of the valley mixing  
induced by the Kekul\'{e} bond ordering.

To understand the revealed robustness of the $n=0$ Landau levels, let us consider the effective low-energy Hamiltonian
with the in-plane Kekul\'{e} distortion, which is given by \cite{VAA} in a basis $({\rm K}_A, {\rm K}_B, {\rm K'}_A, {\rm K'}_B)$ as
$$
 H_{\rm eff} = 
\left(\begin{array}{cc}
\gamma(k_x \sigma_x +k_y\sigma_y) & -i \Delta \sigma_y\\
i\Delta \sigma_y & \gamma(k_x \sigma_x -k_y\sigma_y) \\
\end{array}
\right) = 
\left(\begin{array}{cccc}
0 & \gamma(k_x -i k_y) & 0& - \Delta \\
\gamma(k_x +ik_y) &0&\Delta &0 \\
0& \Delta & 0 &\gamma(k_x  +ik_y) \\
-\Delta & 0 &\gamma(k_x -ik_y) &0
\end{array}
\right) ,
$$
where $\sigma$'s are Pauli matrices, 
$\Delta \equiv 3\Delta_K$ represents the Kekul\'{e}-induced mixing between the K and K' points, and $\gamma=3at/2$ is 
a band parameter.  
In a magnetic field $\bm{B}$, the effective Hamiltonian becomes 
$$
H_{\rm eff}= \left(\begin{array}{cccc}
0 & \gamma \pi & 0& - \Delta \\
\gamma\pi^\dagger &0&\Delta &0 \\
0& \Delta & 0 &\gamma\pi^\dagger \\
-\Delta & 0 &\gamma\pi &0
\end{array}
\right).
$$
Here $\pi = (p_x -i p_y)/\hbar$, where $\bm{p}$ denotes the dynamical momentum $\bm{p} = -i\hbar \bm{\partial} +e \bm{A}$, and $\bm{A}$ stands for the vector potential with $\nabla \times \bm{A} = \bm{B}$. The Hamiltonian  remainss chiral-symmetric, since 
$$
 \Gamma H_{\rm eff} \Gamma = -H_{\rm eff} \quad {\rm with} \quad \Gamma = \mmat{\sigma_z}{0}{0}{\sigma_z}.
$$
For $\Delta =0$, the $n=0$ Landau levels for $H_{\rm eff}$ are exact zero-energy states as given by 
$$
 \psi_0^{\rm K} = \mvecfour{0}{\phi_0}{0}{0} \ {\rm and}\ \psi_0^{\rm K'}= \mvecfour{0}{0}{\phi_0}{0} 
$$
with 
$\pi \phi_0 = 0$.
Note that any linear combinations of those zero-energy eigenstates again belong to the $n=0$ Landau level.

When the Kekul\'{e} distortions are switched on ($\Delta \neq 0$),
the zero-modes are split into two Landau levels with energies as 
$$
 H \psi_\pm = \pm \Delta \psi_\pm .
$$
where
$$
 \psi_+ = \mvecfour{0}{\phi_0}{\phi_0}{0} \ {\rm and}\  \psi_-=\mvecfour{0}{\phi_0}{-\phi_0}{0} 
$$
with  $\pi \phi_0=0$.
As long as the bond disorder is spatially correlated over several lattice constants and hence has little effect on the 
valley mixing, 
its effect must appear in the gauge degrees of freedom alone.  
In such a case, 
the argument due to 
Aharonov and Casher \cite{AC} may be applied in the same way as in the case of $\Delta=0$, which implies 
that those Landau levels are not broadened by disorder.

\section{Zero-modes around topological defects}

Next, we consider a topological defect (vortex) of the Kekul\'{e} bond ordering on the honeycomb lattice, 
where the topological zero-modes around the vortex emerge in the bulk energy gap \cite{HCM,CHJMPS,RMHC,JR} irrespective of the presence or absence of uniform 
magnetic fields perpendicular to the system. 
In Ref.\cite{HFA}, topological interface states localized along 
domain boundaries between different realizations 
of the Kekul\'{e} bond ordering have been discussed.  If we have 
three or more realizations, we can have 
vortices as in Fig.3.  
For vortices, the number of topological modes at $E=0$
is equal to the winding number of the vortex. 
To carry out a quantitative analysis on the zero-modes associated with the topological defects, 
large systems are examined with the kernel polynomial method \cite{KPM} where we can look into 
systems having as large as $10^6$ lattice sites. 
In the implementation, we take the Chebyshev polynomials up to $8200$-th order with the Jackson kernel for the evaluation of the local density of states \cite{KPM}.  

Taking advantage of the large system sizes, we examine the robustness against disorder of the topological zero-modes not only 
for a single vortex with the winding number $n_{\rm w}=1$ but also for the vortices with higher winding numbers $n_{\rm w}>1$. 
We also examine the effect of disorder for the topological zero modes for a pair of a vortex and an anti-vortex. 
In actual numerical calculations, we place the topological structure in the central region of our system
and adopt the open boundary condition at the boundary. As we shall see in the following, the boundary effect is 
negligible as long as we discuss the topological zero-modes localized in the central region of the system.

\subsection{Zero-modes for a single vortex}

\begin{figure}
\includegraphics[scale=0.3]{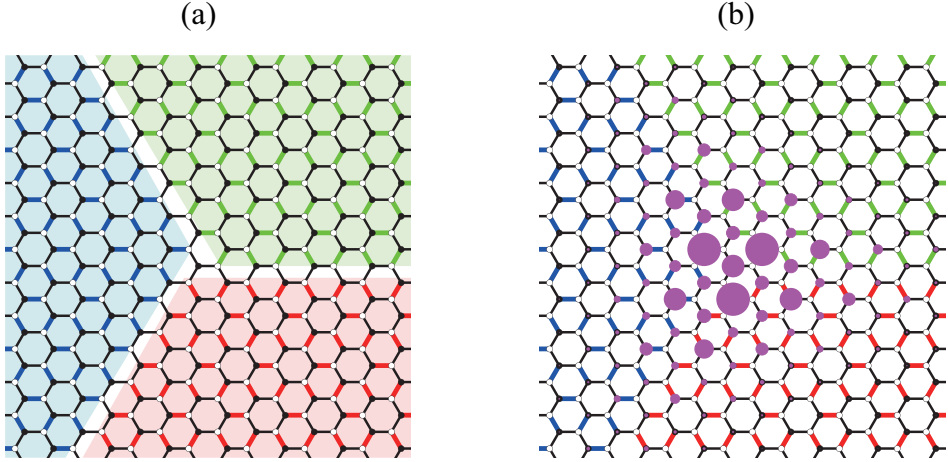}
\caption{
(Color online) (a) A vortex structure of the Kekul\'{e} bond ordering with the winding number $n_{\rm w}=1$. 
Hopping amplitudes with $t+2\Delta_K$ are indicated by thick
blue, green and red bonds, while those with $t-\Delta_K$ 
by thin black bonds. The Kekul\'{e} patterns in the red, 
green, and blue regions are shifted from each other by a single primitive vector of the  original honeycomb lattice. The mismatch  of the three
Kekul\'{e} patterns creates a topological defect, a vortex. The vortex is assumed to be located at the center of the system which has 
$10^6$ sites in the present study. (b) The amplitudes of the zero-mode wave function $\psi(r)$ around   
the vortex, where the radius of each 
purple circle represents $|\psi(r)|$ 
and the zero-mode resides only on the A sub-lattice 
for this vortex. 
The $|\psi(r)|$ satisfies 
$\sum_{r} |\psi(r)|^2 =1$ indicating that the number of zero modes is 1.  
The result shown here is for 
the strength of the Kekul\'{e} ordering 
$\Delta /t =0.15$. 
}
\label{vortex}
\end{figure}

\begin{figure}
\includegraphics[scale=0.7]{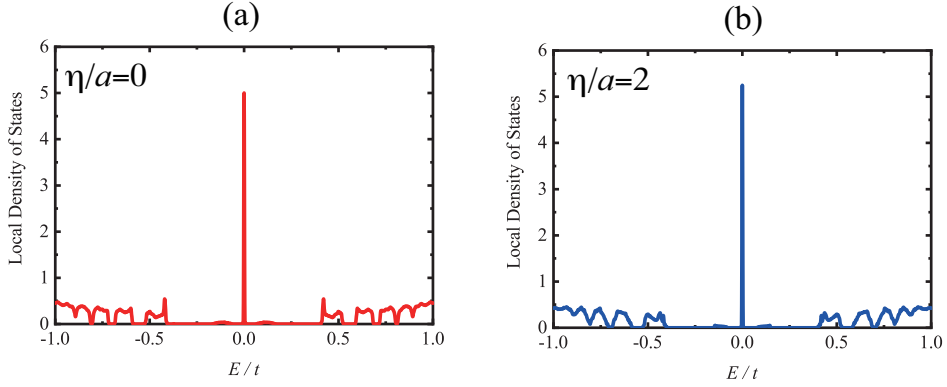}
\caption{
(Color online) The local density of states $\rho(E)$ {\it at the center} of the vortex with $\Delta /t = 0.15$ 
depicted in Fig. \ref{vortex}, evaluated by the 
kernel polynomial method.  The uniform external magnetic field perpendicular to the system is  $\phi/(h/e) =1/50$, 
and the strength of the 
bond disorder is $\sigma/t = 0.1$. An average over 100 realizations of disorder is performed.  
Results for (a) $\eta =0$ (short-ranged) and (b) $\eta/a =2$(long-ranged) are shown. 
}
\label{ldos}
\end{figure}

\begin{figure}
\includegraphics[scale=0.7]{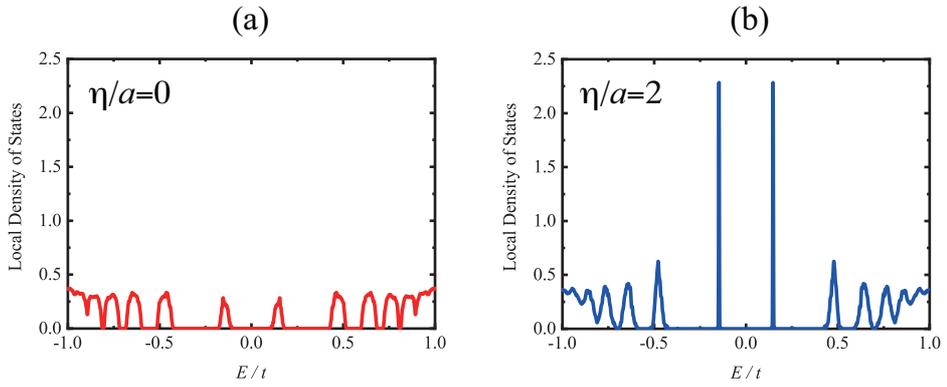}
\caption{
(Color online) The local density of states $\rho(E)$ ($\sim 200$ sites) {\it away form} the center of the vortex 
in Fig. \ref{vortex}, with the parameters the same as in Fig.\ref{ldos}.  
}
\label{ldos_2}
\end{figure}

We first study the robustness of the zero-mode generated by a single vortex with the simplest winding number $n_{\rm w}=1$ 
illustrated in Fig. \ref{vortex}(a). To see the effect of 
spatial correlation of the bond disorder $\delta t_{i,j}$, we again assume that $\delta t_{i,j}$ is Gaussian-distributed with a variance $\sigma$ with the spatial 
correlation length $\eta$ as in the previous section. The numerical results with the kernel polynomial method are shown in Figs \ref{vortex}(b), \ref{ldos} and \ref{ldos_2}.  
Figure \ref{vortex} (b) shows that 
the zero-mode wave function is spatially
localized around  the vortex, and resides only on the A sub-lattice. The local density of states 
at the center site of the vortex in a magnetic field is shown  in Fig. \ref{ldos}.  
It is clearly seen that, irrespective of whether the 
spatial correlation of the bond disorder is long-ranged ($\eta/a =2$)  or short-ranged ($\eta/a =0$), 
the zero-mode associated with the vortex exhibits a sharp peak at $E=0$, suggesting that the 
energy of the zero-mode is unaffected by the bond disorder. The result also suggests that the spatial correlation of disorder is insignificant for the robustness of the zero-modes induced by 
the vortex. We therefore confine ourselves to the spatially uncorrelated disorder for the fermion-vortex 
system in  the following.  To confirm the accuracy of the present numerical approach with the kernel polynomial method, we also evaluate the local density of states well 
away from the vortex center, for example in the red shaded region in Fig. \ref{vortex} (a).  
In such a region,  the effect of the vortex is hardly seen and 
the Landau levels for the uniform Kekul\'{e} bond ordering are recovered [Fig. \ref{ldos_2}] . In particular, the anomalous sharpness of the $n=0$ 
Landau levels is seen to be retained only for 
long-ranged disorder ($\eta /a =2$) in Fig. \ref{ldos_2}(b). This suggests that the present approach
has sufficient accuracy for discussing the robustness of the zero-modes. 


\subsection{Zero-modes for a vortex-antivortex pair}

\begin{figure}
\includegraphics[scale=0.73]{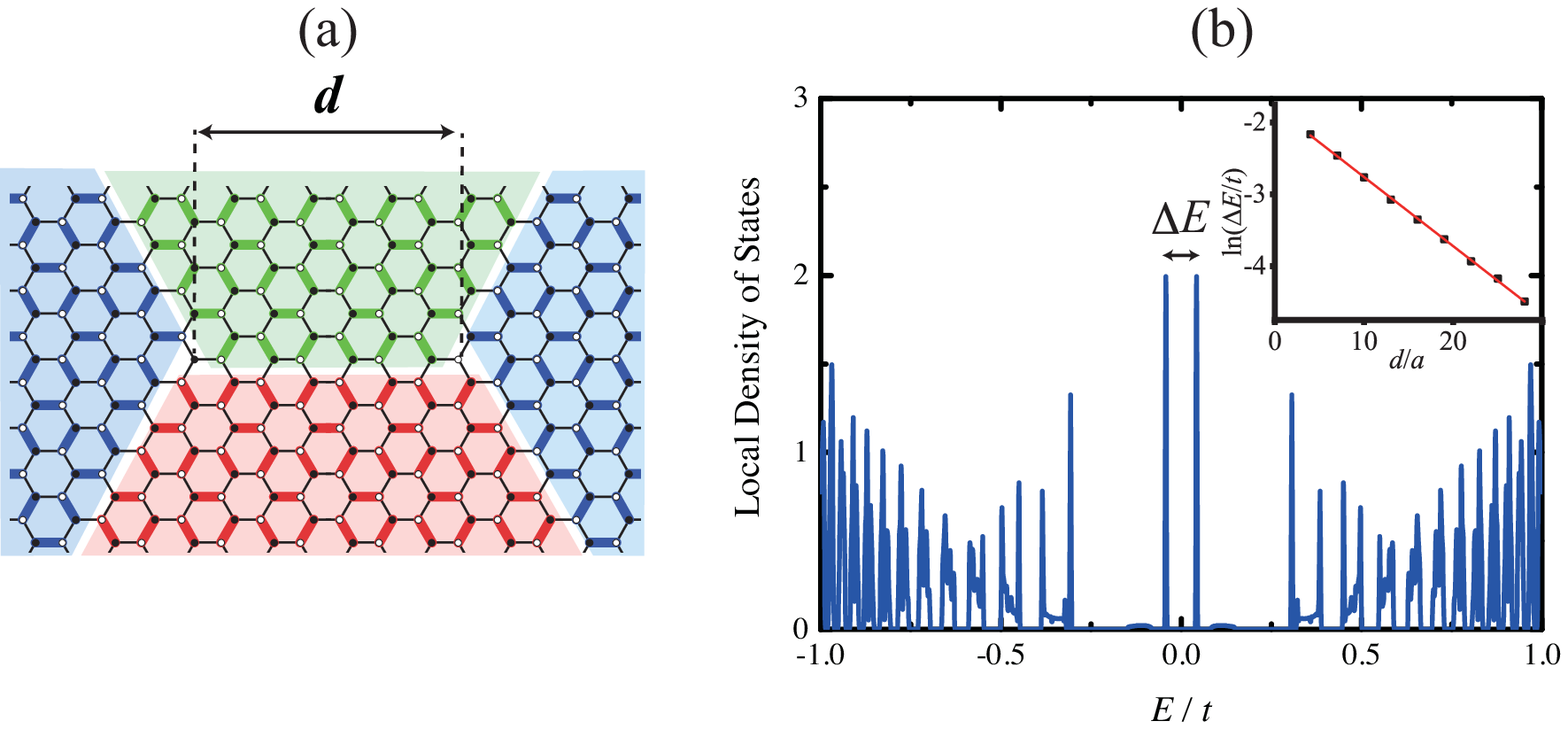}
\caption{
(Color online) (a) A vortex with $n_w=1$ on the left 
and an anti-vortex with $n_w=-1$ on the right 
separated by a distance $d$ in real space. (b) 
The local density of states $\rho(E)$ at the center site (A sub-lattice) of the left vortex for the case of $d/a =4$ in the absence of the bond disorder. 
The system size and the kernel polynomial method are the same as in previous figures. The strength of the Kekul\'{e} bond ordering 
is assumed to be $\Delta /t =0.15$, and the magnetic flux per hexagon is $\phi /(h/e) = 1/100$. 
The inset shows the splitting energy $\Delta E $ 
as a function of the distance $d$ between the vortices. 
}
\label{pair}
\end{figure}

Let us move on to a pair of a vortex with the winding number $n_{\rm w}=+1$ and an anti-vortex with $n_{\rm w}=-1$ as depicted in Fig. \ref{pair} (a).  When the vortex and anti-vortex are close to each other, the 
zero-modes are 
shown in Fig. \ref{pair} (b) to 
split into two. The splitting is expected to be a consequence of the mixing between the vortex zero-mode and the 
anti-vortex zero-mode. 
Indeed, the inset of Fig. \ref{pair}(b) indicates that 
the splitting energy $\Delta E$ decays exponentially with the distance $d$ between the vortices as 
$\Delta E \propto \exp(-d/\xi_0)$. The decay 
length is estimated to be $\xi_0 \simeq 10a$ for the case of 
$\Delta /t =0.15$, which is in good agreement with the localization length of the zero-mode 
wave function given  as $\gamma/\Delta$ \cite{HCM}
for the effective Hamiltonian $H_{\rm eff}$.

\begin{figure}
\includegraphics[scale=0.68]{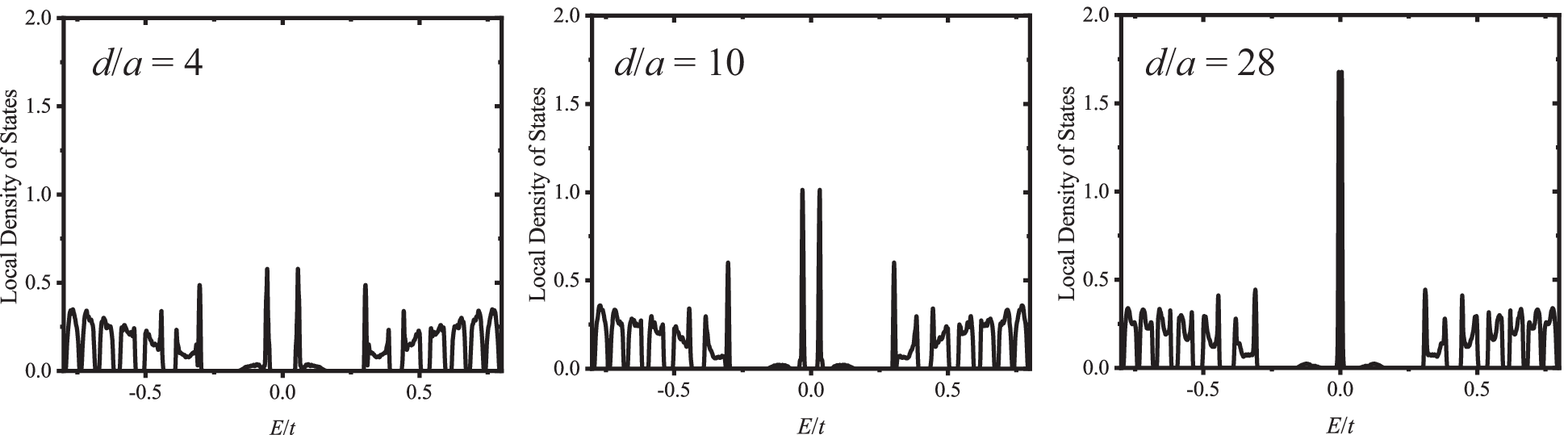}
\caption{In the presence of a bond disorder 
the local density of states $\rho(E)$ evaluated at the center site of the vortex illustrated in  Fig.\ref{pair} (a) 
is shown for the inter-vortex distance $d/a =4$, $10$ and $28$.  The disorder is here assumed to be 
spatially uncorrelated with a uniform distribution 
over $[-W, W]$ with $W/t=0.2$.  
Other parameters are the same as in Fig. \ref{pair}. 
An average over 50 realizations of disorder is preformed. 
}
\label{pair_disorder}
\end{figure}

We then examine the effect of disorder 
by adding a spatially uncorrelated bond disorder. The local density of states at the center site of the vortex 
is shown in Fig. \ref{pair_disorder} for various 
values of the distance between the vortices, $d$. 
We can clearly see that the disorder degrades the sharpness of the split zero-modes when the vortices are 
close to each other. 
The sharpness is recovered as $d$ is increased, 
where the splitting becomes small.  
The result indicates that the distance between vortex and anti-vortex 
has to be larger than the localization length of the zero-mode for the robustness of the zero-mode. 

\subsection{Zero-modes for vortices having higher winding numbers}

\begin{figure}
\includegraphics[scale=0.75]{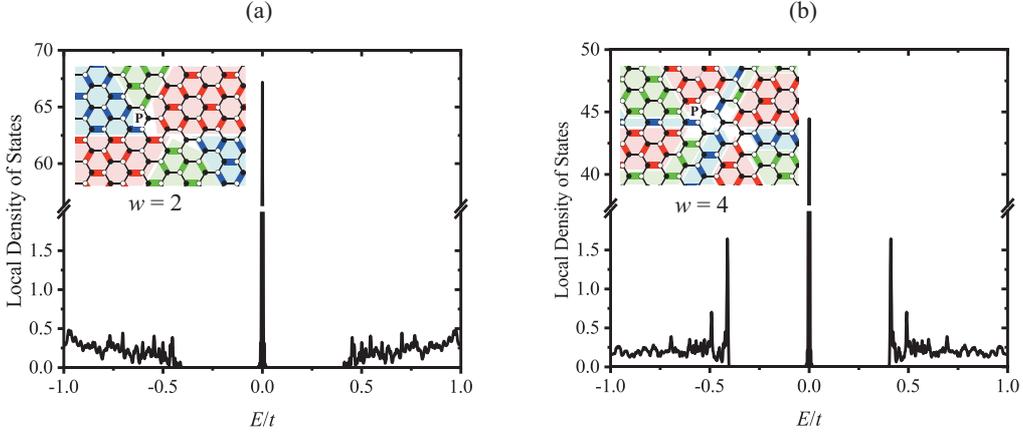}
\caption{
The local density of states $\rho(E)$ at the site P 
($\in$ A sub-lattice) located in the central region of the vortex with (a) the winding number $n_{\rm w}=2$, and (b) $n_{\rm w}=4$ as 
illustrated respectively in each inset, in zero 
magnetic field. The strength of the Kekul\'{e} bond ordering is $\Delta /t =0.6$, and 
a bond disorder, spatially uncorrelated, is considered 
with a uniform distribution over $[-W, W]$ with $W/t=0.2$. Here a result for a single realization of disorder is presented. 
}
\label{w2_disorder}
\end{figure}

So far we have considered the simplest winding 
number $n_{\rm w}=1$.  How about the vortices having 
higher winding numbers?  
So let us look at the result with the winding 
number for $n_{\rm w}=2$ [Fig. \ref{w2_disorder}(a)] 
and $n_{\rm w}=4$ [Fig. \ref{w2_disorder}(b)], 
in zero 
magnetic field to single out the effect of $n_w$. 
The local density of states in the central region 
of the vortex evaluated by the kernel polynomial method 
is shown in the figure.   
We can clearly see that the energy of the zero-modes is unaffected by the bond disorder for these higher 
winding numbers. By evaluating he total amplitude 
around the vortex, we confirm that the number of zero modes at $E=0$ is indeed equal to the winding number of the vortex. This indicates 
that all the zero modes for the vortex with a winding number greater than 1 are degenerated at $E=0$ even in the presence of the bond disorder.  
To endorse the topological origin of the zero-modes, we further examined a system with the same winding number but having different bond  patterns 
in the central region of  the vortex. 
We then find that the  zero-modes are 
insensitive to the detailed structure of the central region of the vortex, which suggests that the emergence of the 
zero-modes at the vortex is indeed 
dictated by the non-local topological property, such as the winding number, of the vortex structure.
All the zero-modes again reside only on the A sub-lattice, and are still the eigenstates of the chiral operator with the eigenvalue 1. 

\subsection{Staggered potential and irrational charges}

What will happen when the chiral symmetry 
is broken, typically by a staggered potential defined as the Hamiltonian,
$$
 H_\mu = \mu \left(\sum_{i \in A} c_i^\dagger c_i - \sum_{i\in B}c^\dagger_ic_i\right).
$$ 
When the chiral symmetry is degraded, 
irrational charges associated with vortices have 
been discussed \cite{HCM,CHJMPS,RMHC}. Here we can examine 
whether the irrational charges remain robust against the bond disorder with direct numerical calculations. 
Even in the presence of the bond disorder which respects the chiral symmetry, the energy of the zero-mode 
associated with the vortex must be exactly $E=\mu$, since the 
zero-mode has amplitudes only on the A sub-lattice, 
hence an eigenstate of $H_\mu$. This is a direct consequence of the chiral symmetry for  $\mu=0$ that 
can be confirmed by numerical calculations.

\begin{figure}
\includegraphics[scale=0.3]{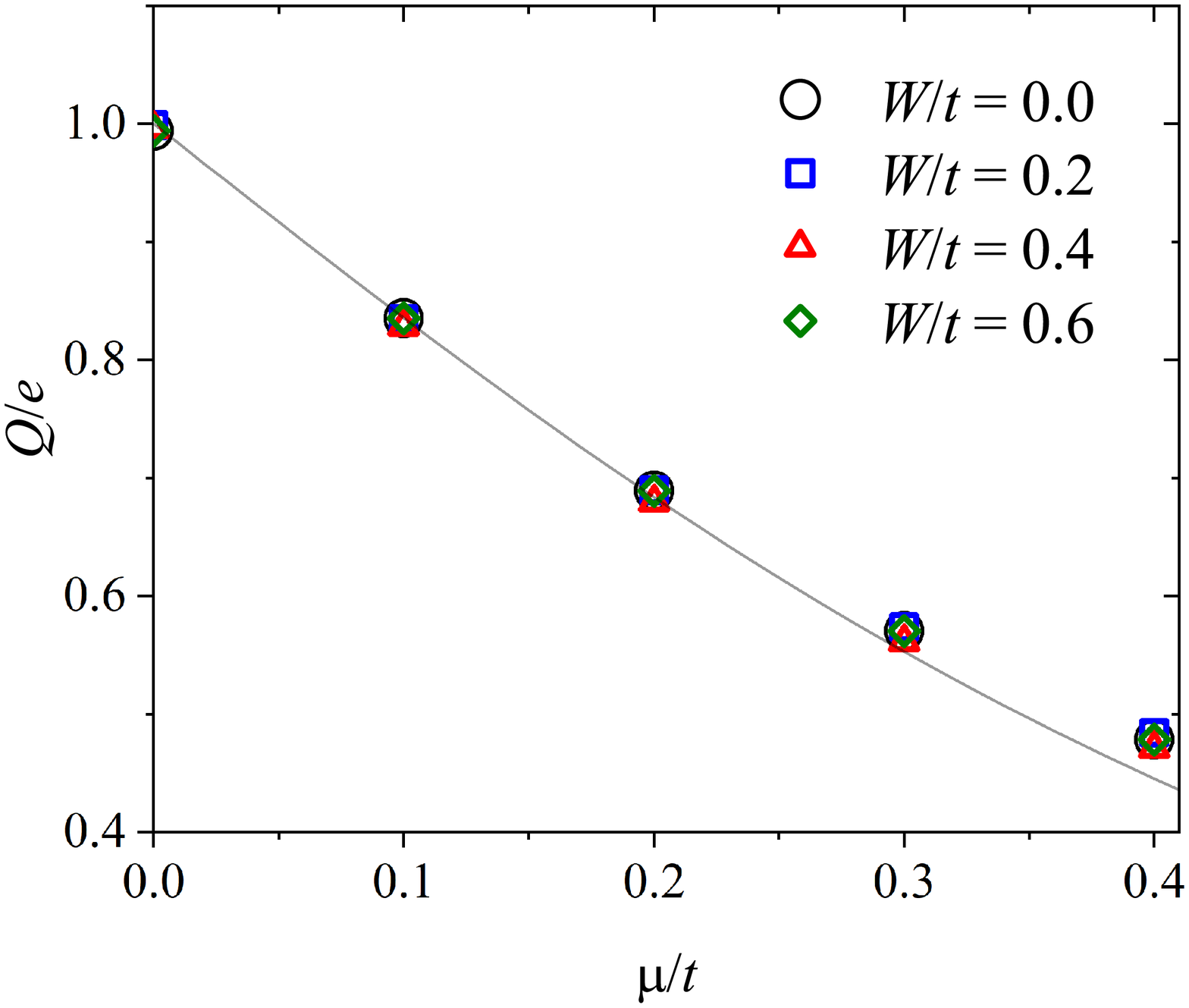}
\caption{
The fractional charge, $Q$, associated with the vortex with winging number $n_w=2$ plotted against the 
staggered potential $\mu$ for various values of the 
strength of disorder $W$. 
Spatially uncorrelated bond disorder with a uniform distribution 
in the range $[-W/2, W/2]$ is adopted. The curve represents  the formula $(n_w/2)(1-\mu/\sqrt{\mu^2+\Delta^2})$ in Ref. \cite{CHJMPS}.
}
\label{irrational}
\end{figure}

The charge 
of a vortex is defined as the increment in the local charge when the vortex is introduced \cite{HCM}. 
We evaluate the charge 
for various values of the disorder strength and the staggered potential $\mu$ for the vortices with $w=1 - 4$.  
The result indicate 
that the charge is proportional to the winding number $n_{\rm w}$, and insensitive to the strength of the bond disorder. 
This is exemplified for $n_w=2$ 
for various values of the disorder $W$ in Fig. \ref{irrational}. The insensitivity of the irrational charge to disorder may be related to the fact that 
the bond disorder does not change the energy $E= \mu$ of the zero modes.  

\section{Summary}

We have investigated numerically the honeycomb lattice with the Kekul\'{e} bond ordering to examine the 
robustness of zero-modes against disorder that respects the chiral symmetry. 
We have clearly shown the following: 
(i) The split $n=0$ Landau levels 
retain the anomalous sharpness as soon as the 
bond disorder becomes spatially correlated over several lattice constants, which implies that the valley mixing 
arising from the 
Kekul\'{e} ordering does not impair the sharpness of the $n=0$ Landau levels.  
(ii) We have analyzed  the topological zero-modes induced by the vortex in the  Kekul\'{e} bond ordering.
With the kernel polynomial method for 
large systems, 
the zero-modes induced by 
the vortex turns out to be robust against the bond disorder irrespective of the presence or the absence of the spatial correlation, confirming 
the topological stability of the zero-energy states.  
(iii) We have also analyzed the electronic states around the vortices with higher winding numbers as well as a 
vortex-antivortex pair.  
(iv) The fractional and irrational charge is evaluated numerically for vortices with vairous winding numbers. It is demonstrated that the charge 
is proportional to the winding number of vortex, in reasonable agreement with the effective theory \cite{CHJMPS}. 
The fractional charge is insensitive to the bond disorder 
that espects the chiral symmetry. 
It is an interesting future problem how we can 
relate these with experimentally observable properties.

\section*{Acknowledgements}
This work is partly supported by JSPS KAKENHI Grants Numbers JP19K03660 (TK), and JP17H06138.





\end{document}